\documentclass[nofootinbib,preprintnumbers,prd,superscriptaddress,twocolumn,showpacs]{revtex4}

\usepackage{amsmath}
\usepackage{amsfonts}
\usepackage{amssymb}
\usepackage{graphicx}
\usepackage[titletoc]{appendix}
\usepackage{hyperref}
\usepackage{cleveref}
\usepackage[rightcaption]{sidecap}
\usepackage{subfigure}
\usepackage{comment}

\usepackage{dcolumn}

\usepackage{array}
\usepackage{ctable}
\usepackage{multirow}
\usepackage{siunitx}
\usepackage{longtable}
\usepackage{tabularx}
\usepackage{booktabs}

\usepackage{amsthm}
\usepackage{mathrsfs}

\usepackage{epsfig}
\usepackage{amscd}

\usepackage{bm}
\usepackage{natbib}
\usepackage{url}
\usepackage{xspace}
\usepackage{kantlipsum}

\graphicspath{{Graphics/}}

\def\be{\begin{equation}}
\def\ee{\end{equation}}
\def\bea{\begin{eqnarray}}
\def\eea{\end{eqnarray}}

\def\prl{Phys. Rev. Lett.}

\def\apj{ApJ}
\def\apjl{ApJ Lett.}

\definecolor{vividviolet}{rgb}{0.62, 0.0, 1.0}
\definecolor{amaranth}{rgb}{0.9, 0.17, 0.31}
\definecolor{palatinateblue}{rgb}{0.15, 0.23, 0.89}
\definecolor{brightpink}{rgb}{1.0, 0.0, 0.5}
\definecolor{cornflowerblue}{rgb}{0.39, 0.58, 0.93}
\definecolor{deepcarminepink}{rgb}{0.94, 0.19, 0.22}
\definecolor{radicalred}{rgb}{1.0, 0.21, 0.37}
\hypersetup{ linktoc=all,
    colorlinks, linkcolor={palatinateblue},
    citecolor={brightpink}, urlcolor={amaranth}
}

\begin{document}

\title{Repulsive gravity in regular black holes}

\author{Orlando~Luongo}
\email{orlando.luongo@unicam.it}
\affiliation{Universit\`a di Camerino, Via Madonna delle Carceri 9, 62032 Camerino, Italy.}
\affiliation{SUNY Polytechnic Institute, 13502 Utica, New York, USA.}
\affiliation{Istituto Nazionale di Fisica Nucleare, Sezione di Perugia, 06123, Perugia,  Italy.}
\affiliation{INAF - Osservatorio Astronomico di Brera, Milano, Italy.}
\affiliation{NNLOT, Al-Farabi Kazakh National University, Al-Farabi av. 71, 050040 Almaty, Kazakhstan.}

\author{Hernando Quevedo}
\email{quevedo@nucleares.unam.mx}
\affiliation{Instituto de Ciencias Nucleares, Universidad Nacional Aut\'onoma de M\'exico, Mexico.}
\affiliation{Dipartimento di Fisica and Icra, Universit\`a di Roma “La Sapienza”, Roma, Italy.}

\begin{abstract}
We evaluate the effects of repulsive gravity using first order geometric invariants, \textit{i.e.}, the Ricci scalar and the eigenvalues of the Riemann curvature tensor, for three regular black holes, namely the Bardeen, Hayward, and Dymnikova spacetimes. To examine the repulsive effects, we calculate their respective onsets and regions of repulsive gravity. Afterwards, we compare the repulsive regions obtained from these metrics among themselves and then with the predictions got from the
Reissner-Nordstr\"{o}m and Schwarzschild-de Sitter. We find that the Dymnikova spacetime does not exhibit regions in which gravity changes its sign. A notable characteristic, observed in all these metrics, is that the repulsive regions appear to be unaffected by the mass that generates the regular black hole. This property emerges due to the invariants employed in our analysis, which do not change sign through \emph{linear combinations} of the mass and the free coefficients of the metrics. As a result, gravity can change sign independently of the specific values acquired by the mass. This conclusion suggests a potential \emph{incompleteness} of regular solutions, particularly in terms of their repulsive effects. To further highlight this finding, we numerically compute, for the Reissner-Nordstr\"{o}m and Schwarzschild-de Sitter solutions, the values of mass, $M$, that  emulate the repulsive effects found in the Bardeen and Hayward spacetimes. These selected values of $M$ provide evidence that regular black holes do not incorporate  repulsive effects  by means of the masses used to generate the solutions themselves. Implications and physical consequences of these results are then discussed in detail.
\end{abstract}


\maketitle
\tableofcontents

\section{Introduction}

Black holes (BHs) recently provided new insights into the understanding of our universe, particularly with modern discoveries of gravitational waves \cite{2016PhRvL.116f1102A} and BH shadows \cite{2019ApJ...875L...1E}. These objects hold great significance as they exist within intense gravitational regimes, potentially resulting in deviations from Einstein's theory of gravity. These departures may result in quantum gravity effects that cannot be avoided at the fundamental level, leading to expected breakdowns of gravity\footnote{Classically, Penrose and Hawking established theorems \cite{1965PhRvL..14...57P, 1970RSPSA.314..529H} which show that matter presence, satisfying plausible energy conditions, leads inevitably to singularities. These singularities, in highly gravity regimes, may be avoided by quantum gravity \cite{avoid}.} \cite{Astashenok:2014nua,Astashenok:2013vza,Astashenok:2017dpo,Capozziello:2019cav,Astashenok:2020qds}.

In addition to this scenario, naked singularities have been shown to be conceivable possibilities under quite general assumptions \cite{primaref1,primaref2}. Specifically, BH solution may possess a corresponding naked singularity counterpart\footnote{It is worth noting that, under certain circumstances, naked singularities can arise throughout the evolution of mass distributions in a gravitational collapse \cite{Luongo:2014qoa}.}. Both BHs and naked singularities suggest the possibility of repulsive regions, where gravity exhibits a change of sign.

It is therefore licit to expect that repulsive regions may also occur in further contexts and, in particular, in alternative typologies of solutions of Einstein's field equations. Among all the possibilities, recently a renewed interest in the physics of regular BHs arose, as they can model compact objects in a similar manner than singular solutions, see e.g. \cite{Boshkayev:2023rhr,Boshkayev:2022haj}, appearing \emph{de facto} as robust physical scenarios\footnote{Further, approaches that suggest geometric quasi-particles have been studied. In such scenarios, it appears natural to presume regular solutions as possible metrics that classically describe those particles, see e.g. \cite{Belfiglio:2022cnd,Belfiglio:2022egm,Belfiglio:2022yvs,Fedichev:2003bv}.}.

However, in all these scenarios, including BHs, regular solutions and/or naked singularities, finding out  regions where gravity inverts its sign represents an arduous challenge since it is clearly needful to formulate an \emph{invariant strategy} to single out the regions where gravity modifies its character\footnote{For the sake of completeness, the effects of repulsive gravity are also present in cosmological scenarios \cite{Copeland:2006wr,Aviles:2012ay,Capozziello:2019cav}. It is possible to account for them, through the use of barotropic fluids \cite{Capozziello:2017buj,Dunsby:2016lkw,Aviles:2014mua}, or more broadly by virtue of scalar fields \cite{Luongo:2014nld,Boshkayev:2019qcx}. Recently, it has been proposed that these effects could be imputed to mechanisms of vacuum energy destruction  \cite{Luongo:2018lgy,DAgostino:2022fcx,Belfiglio:2022egm} or production from BHs \cite{2023ApJ...943..133F,2023arXiv230409817M,Parnovsky:2023wkc}.}.  In this respect, it turns out that standard approaches used to define repulsive gravity, such as null cones and effective potentials, are \emph{frame-dependent} \cite{defelice}. To define repulsive gravity in an invariant manner, one could use curvature invariants. However, quadratic invariants, such as the Kretschmann  and the Chern-Pontryagin scalars, cannot reproduce the behavior of the simplest naked singularities described by the Schwarzschild metric with negative mass.

Recently, we proposed in Refs. \cite{mymodel1,Luongo:2014qoa} an alternative approach that is based upon the use of first-order curvature invariants, namely, the \emph{eigenvalues of the Riemann curvature tensor}.

In this work, we will use this invariant approach and will compare our results with the behavior of the Ricci scalar, which is non-zero in the case of regular BHs. To do so, we study the repulsive behavior of the above-cited regular BHs. These represent smooth  asymptotic flat frameworks, where a   non-singular center appears \cite{bardeen1968proceedings}. Thus, we focus on spherical and static spacetimes and limit our analysis to the most popular regular BHs, \textit{i.e.}, the Bardeen, Hayward, and Dymnikova solutions \cite{1992GReGr..24..235D}, each of them exhibiting different physical interpretations. Specifically, Bardeen metric possesses a topological charge, Hayward metric includes a de Sitter regular contribution, and the Dymnikova spacetime is characterized by a non-linear electrodynamic source. We calculate the curvature eigenvalues for the above spacetimes, which are first-order invariants,  leading \emph{de facto} to an invariant determination of the corresponding  repulsive zones, where gravity changes its sign.

Further, we also compare our findings from each solution among them and with the Reissner-Nordstr\"om and Schwarzschild-de Sitter spacetimes. We underline that repulsive regions can shed light on the physical nature of regular BHs. Indeed, we remark that the repulsive radii do not depend on the mass values of the corresponding spacetimes.  In other words, if the regular solutions are applied to describe compact objects, it is expected that repulsive gravity would depend on a combination of the free parameters entering each solution. As this does not occur, we conclude that \emph{either these metrics are incomplete or
the mass cannot be considered as a source of repulsive gravity as in the case of singular BHs} \cite{Luongo:2014qoa}.
All these scenarios are thoroughly discussed and critically reviewed, showing the numerical values of masses for Schwarzschild-de Sitter and Reissner-Nordstr\"{o}m metrics that  can mimic the properties of regular solutions.

The paper is organized as follows. In Sec.~\ref{sezione2}, repulsive gravity is discussed. In Sec.~\ref{sezione3}, the employed regular BHs are introduced and the repulsive effects calculated. In Sec.~\ref{sezione4}, the theoretical consequences of our findings are reported. There, we also compare our results with the outcomes in the case of the Reissner-Nordstr\"om and Schwarzschild-de Sitter spacetimes.  Finally, in Sec.~\ref{sezione5}, conclusions and perspectives are reported\footnote{Throughout this paper we use natural units, $G=c= \hbar = 1$, and Lorentzian  signature $(-,+,+,+)$.}.


\section{Invariant approach to repulsive gravity}\label{sezione2}

Repulsive gravity is a potential phenomenon that is not ruled out by theoretical frameworks in high-gravity regimes. In analogy to gravity, where an invariant definition is allowed through spacetime curvature invariants, it is conceivable that repulsive effects could be characterized by invariant quantities. The corresponding \emph{gravitational charge}  is  mass, $M$. Thus, one may expect that any invariant representation of repulsive gravity might exhibit linear mass terms, \textit{i.e.}, those explicitly showing the mass sign (positive or negative for attractive and repulsive gravity, respectively). Indeed, if the change of sign, $M\rightarrow -M$, occurs, then repulsion regions are possible. This can be seen in the behavior of the Ricci scalar, $R$, in the case of non-vacuum solutions.
Quadratic invariants would be proportional to $M^2$, hiding the sign of gravity \cite{mymodel1}.

As a suitable prescription to overcome the aforementioned issue, it is  possible to employ the
curvature eigenvalues, which are linear in the components of the curvature.

The corresponding method is general and appears valid for any  spacetimes. To do so, let us start by noticing that repulsive gravity definition can be established using an \emph{orthonormal frame} whose advantages are

\begin{itemize}
    \item[-] they allow observers to conduct local measurements of time, space, and gravity;
    \item[-] all quantities associated with this frame are independent of coordinates.
\end{itemize}

The frame is represented by the tetrads $\vartheta^a$, defined through differential forms, which for a given  line element  $ds^2 = g_{\mu\nu} dx^\mu dx^\nu$, read
\begin{equation}\label{fn}
ds^2 =  \eta_{ab}\vartheta^a\otimes\vartheta^b\,,
\end{equation}
\noindent with $\eta_{ab}={\rm diag}(-1,1,1,1)$ and $\vartheta^a = e^a_{\ \mu}dx^\mu$. Thus, adopting the Cartan formalism that determines the connection one-form $\omega^a_{\ b}$ and the curvature two-form $\Omega^a_{\ b}$ by means of the structure equations
\begin{subequations}
\begin{align}
d\vartheta^a &=- \omega^a_{\ b }\wedge \vartheta^b\,,\\
\Omega^a_{\ b} &= d\omega^a_{\ b} + \omega^a_{ \ c} \wedge \omega^c_{\ b} = \frac{1}{2} R^a_{\ bcd} \vartheta^c\wedge\vartheta^d\,,
\end{align}
\end{subequations}
the tetrad components of the Riemann tensor $R^a_{\ bcd}$ appear as the coefficients of the decomposition of the curvature two-form $\Omega^a_{\ b}$ in terms of the local tetrad $\vartheta^a$.

\subsection{The bivector representation}

There are several methods to compute the eigenvalues of the curvature tensor \cite{stephani}. When using an orthonormal local tetrad, the simplest method consists in representing the components of the Riemann tensor in terms of local bivectors \cite{mtw}. To this end, we introduce the bivector index  $A= 1, \ldots, 6$, which corresponds to two tetrad indices, $A \rightarrow ab$, leading to the representation of ${\bf R}_{AB} \rightarrow R_{abcd}$. For concreteness, we choose the bivector indices $A \rightarrow ab$ according to the following convention
\begin{displaymath}
1 \rightarrow 01,\ 2 \rightarrow 02,\ 3 \rightarrow 03,\ 4 \rightarrow 23, \ 5 \rightarrow 31, \ 6 \rightarrow 12.
\end{displaymath}
Using this notation, the curvature tensor can be expressed as the  $(6 \times 6)$-matrix $R_{AB}$ which, using Einstein's field equations, can be written as \cite{ory}

\be \label{eq: CurvatureTensor}
{\bf R}_{AB}=\left(
\begin{array}{cc}
	{\bf M}_1 & {\bf L} \\
	{\bf L} & {\bf M}_2 \\
\end{array}
\right),
\ee
where the explicit form of the $(3\times 3)$-matrices ${\bf M_1}$, ${\bf M_2}$, and ${\bf L}$ is reported  in detail
in the Appendix.

According to recent developments \cite{ory,mymodel1,Luongo:2014qoa,Giambo:2020jjo,Luongo:2015zaa}, one can  identify  repulsive gravity regions by assuming that a change in the gravitational field implies the  eigenvalue sign to change as well, which is a consequence of assuming that curvature is a  measure of the gravitational interaction. Correspondingly, we emphasize that
\begin{itemize}
    \item[-] by virtue of asymptotic flatness, eigenvalues vanish at spatial infinity as well as the gravitational field of compact sources,  among which regular BHs constitute a particular case;
    \item[-] as the central source is approached, and in the presence of attractive gravity only, the eigenvalues tend to increase monotonically;
    \item[-] if repulsion is dominant, a change in the sign of \emph{at least one eigenvalue} should occur;
    \item[-] the location at which the passage from attractive to repulsive gravity occurs is uniquely determined by the zeros of the eigenvalues.
\end{itemize}

In view of the above, the location where an  eigenvalue vanishes characterizes entirely the point at which the sign of gravity  changes. Consequently, in the case of spherically symmetric spacetimes, we define the radius of repulsion,  $r_{rep}$, as the \emph{first extremal} occurring in a curvature eigenvalue, when the
source of gravity is approached from infinity.

In addition, this method to find out repulsion regions can be simply adapted to regular BHs since they provide  properties similar to those of BHs, albeit behaving smoothly at the origin. Clearly, this characteristic does not influence  the use of curvature eigenvalues and so, baptizing the eigenvalues as $\lambda_i,\ i=1,\ldots 6$, we compute them for the matrix ${\bf R}_{AB}$.

Consequently, we can assume that the  eigenvalues

\begin{itemize}
\item[-] contain \emph{all the information} about curvature;
    \item[-]  behave as scalars under coordinate transformations;
    \item[-] the change of sign of at least one  eigenvalue indicates the transition to repulsive gravity;
    \item[-] the presence of an extremal in an eigenvalue indicates the onset of repulsive gravity;
    \item[-] the zeros of the eigenvalues indicate the locations where repulsive gravity becomes dominant.
\end{itemize}

Furthermore, in the case of spherically symmetric spacetimes, the repulsion radii, $r_{rep}$, that determine the onset  repulsive gravity are determined by the condition
\begin{equation}\label{derlam}
    \frac{\partial \lambda_i}{\partial r}=0\, .
\end{equation}

In what follows, we search for these regions in  regular BH solutions.

\begingroup
\squeezetable
\begin{table*}
\caption{Table of comparison among eigenvalues for different spacetimes. We  compare $\lambda_i^{B;H}$ with the same subscript for every metric, namely Bardeen, Hayward with the Reissner-Nordstr\"{o}m and Schwarzschild-de Sitter eigenvalues. We compute the value of mass that permits to obtain the exact matching between the two radii, namely the radii allowing the eigenvalues to vanish. Hence, here the values of the masses for the singular solutions that permit to emulate the results of regular BHs are reported. In this respect, we notice that repulsive gravity occurs only at a precise value of mass. Hence, regular solutions appear less general, in terms of repulsive effects, than standard BHs. The presence of opposite radii implies that the sign of $\Lambda$ and $q$ might be opposite (discord) and not well-defined. Finally, the regular solution clearly can be mapped into singular ones, at the level of repulsive gravity, by evaluating the precise values of masses reported below.}
\begin{tabular}{@{}ccccccccccccccccccccccccc@{}}
 \toprule
& \\[-6pt]
\hline
\hline
\cmidrule{5-15} \cmidrule{15-24}
& &&& \multicolumn{1}{c}{$M^{(RN)}$} &&&&&&\multicolumn{7}{c}{$\qquad\,\, M^{(SdS)}$}\\
\cmidrule{6-7} \cmidrule{7-11}
& $\#$   & $\lambda_1^{RN}$ &\phantom{a}& $\lambda_2^{RN}$ &\phantom{a}& $\lambda_3^{RN}$
                    &\phantom{abcdef} & $\lambda_1^{SdS}$ &\phantom{a}& $\lambda_2^{SdS}$ &\phantom{a}& $\lambda_3^{SdS}$ &\phantom{a}   \\
\midrule
\hline
\emph{Bardeen}:\\
 &$\lambda_1^B$ & $0.65q\wedge3.46Q$ && --  && $0.43q\wedge 2.30q$  && $ -0.04q^3 \Lambda\wedge-6.12q^3 \Lambda$ && $ 0.04q^3 \Lambda\wedge6.12q^3 \Lambda$  && $ 0.04q^3 \Lambda\wedge6.12q^3 \Lambda$ &\\

 &$\lambda_2^B$ & $1.06q$ && $1.06q$  && $0.71q$  && $-1.41q^3\Lambda$ && $1.41 q^3 \Lambda$  && $1.41 q^3 \Lambda$ & \\[4pt]
\hline
\emph{Hayward}:\\
 &$\lambda_1^H$ & $2.26q^2a^{-2/3}$ && $-2.26q^2a^{-2/3}$  && $1.51q^2a^{-2/3}$  && $-0.15 a^2 \Lambda$ && $-0.15 a^2 \Lambda$  && $0.29 a^2 \Lambda$ &\\

 &$\lambda_2^H$ & $0.42q^2a^{-2/3}$ && $-0.42q^2a^{-2/3}$  && $0.63q^2a^{-2/3}$  && $2a^2\Lambda$ && $-2a^2\Lambda$  && $0.29 a^2 \Lambda$ & \\[4pt]
 \bottomrule
 \hline
 \hline
\end{tabular}
\label{Table}
\end{table*}
\endgroup

\section{Static spherically symmetric regular black holes}\label{sezione3}

The  basic strategy to describe compact objects is by using a spherically symmetric and non-rotating metric. The spacetime geometry can be expressed as

\begin{equation}
ds^2 = -f(r)dt^2 + \frac{1}{f(r)}dr^2 + r^2(d\theta^2 + \sin^2\theta d\phi^2)\,,
\end{equation}
which can also be used to study electrically charged objects in general relativity as well as in extended theories of gravity. In this work, we focus on the Bardeen, Hayward, and Dymnikova regular spacetimes. These metrics have simple structures that are well-suited for our analysis of repulsive gravity.

\subsection{Repulsive gravity in Bardeen spacetime}

The Bardeen metric \cite{bardeen1968proceedings,2000PhLB..493..149A} represents a non-rotating BH, solution of Einstein's field equations, exhibiting topological charge, first derived with the aim of finding a solution that describes a magnetically-charged BH, alternative to the traditional Reissner-Nordstr\"{o}m metric. The lapse function is
\begin{equation}
f (r) =  1 - \frac{2 M r^2}{(r^2 + q^2)^{3/2}}\,,
\end{equation}
where  $q$ and $M$ represent the magnetic or topological charge  and the magnetic monopole mass, respectively, while the Schwarzschild BH is recovered once $q$ tends to zero.

The corresponding solution is reinterpreted noticing that, using the Newman-Janis algorithm, the Bardeen metric is equivalent to
the Kerr spacetime\footnote{This also indicates that rotating regular BHs may be, in general, generated through the Newman-Janis algorithm.} \cite{Bambi:2014nta}.

Computing the curvature eigenvalues, $\lambda_i^{(B)}$, for this metric, we  can write
\begin{subequations}\label{lambdabardeen}
\begin{align}
\lambda_1^{(B)} & = - M\frac{ 2q^4 - 11 q^2 r^2 + 2 r ^4}{(q^2 + r^2)^{7/2}}\,, \\
\lambda_2^{(B)} & =  \lambda_3^{(B)} = - \lambda_5^{(B)} = - \lambda_6^{(B)} =  - M\frac{ 2q^2 - r ^2}{(q^2 +  r^2)^{5/2}}\,, \\
\lambda_4^{(B)} & = \frac{2M}{(q^2 + r^2)^{3/2}}\,.
\end{align}
\end{subequations}

On the other side, evidence for repulsive gravity could be inferred also from the Ricci scalar, which is a first order curvature invariant. In this case, we obtain
\begin{equation}\label{riccibardeen}
R^{(B)}=-6M\frac{q^2\left( r^2-4\,q^2\right)}{\left(r^2+q^2\right)^{7/2}}\,.
\end{equation}
From Eqs. \eqref{lambdabardeen}, we see that $\lambda_1$ and $\lambda_2$ change their sign when approaching $r=0$, indicating the places at which repulsive gravity becomes dominant.

The first extremum that is reached when approaching from infinity is located at $r^2 =(21+\sqrt{345}) q^2 /4$, which corresponds to a local minimum of $\lambda^{(B)}_2$, whereas a maximum value is located at  $r^2 =(21-\sqrt{345}) q^2 /4$. On the other hand, $\lambda^{(B)}_2$ shows a maximum at $r=2q$ and changes its sign at $r= \sqrt{2}q$.

Accordingly, the largest extremum belongs to $\lambda_1^{(B)}$ and determines the repulsion radius as $r_{rep}\approx 3.14 q$, corresponding to the onset of repulsive gravity. Moreover, when approaching the source from infinity, the first zero is in $\lambda_1^{(B)}$ at $r=r_{dom}=2.3 q$, which is the point at which repulsive gravity becomes dominant.

Moreover, the Ricci scalar for the Bardeen metric changes sign at a given radius, say $r_{R,B}^\star$, as
\begin{equation}
r_{R,B}^\star=2q\,,
\end{equation}
exhibiting repulsive gravity at distinct domains. Indeed, since $r$ is non-negative and $q$ positive or negative, it appears evident from Eq. \eqref{riccibardeen} that repulsive gravity occurs at $r>2q$ as $q$ is positive, while at $0<r<2|q|$ for negative charges.

\begin{figure*}
 \centering
\includegraphics[width=1\columnwidth,clip]{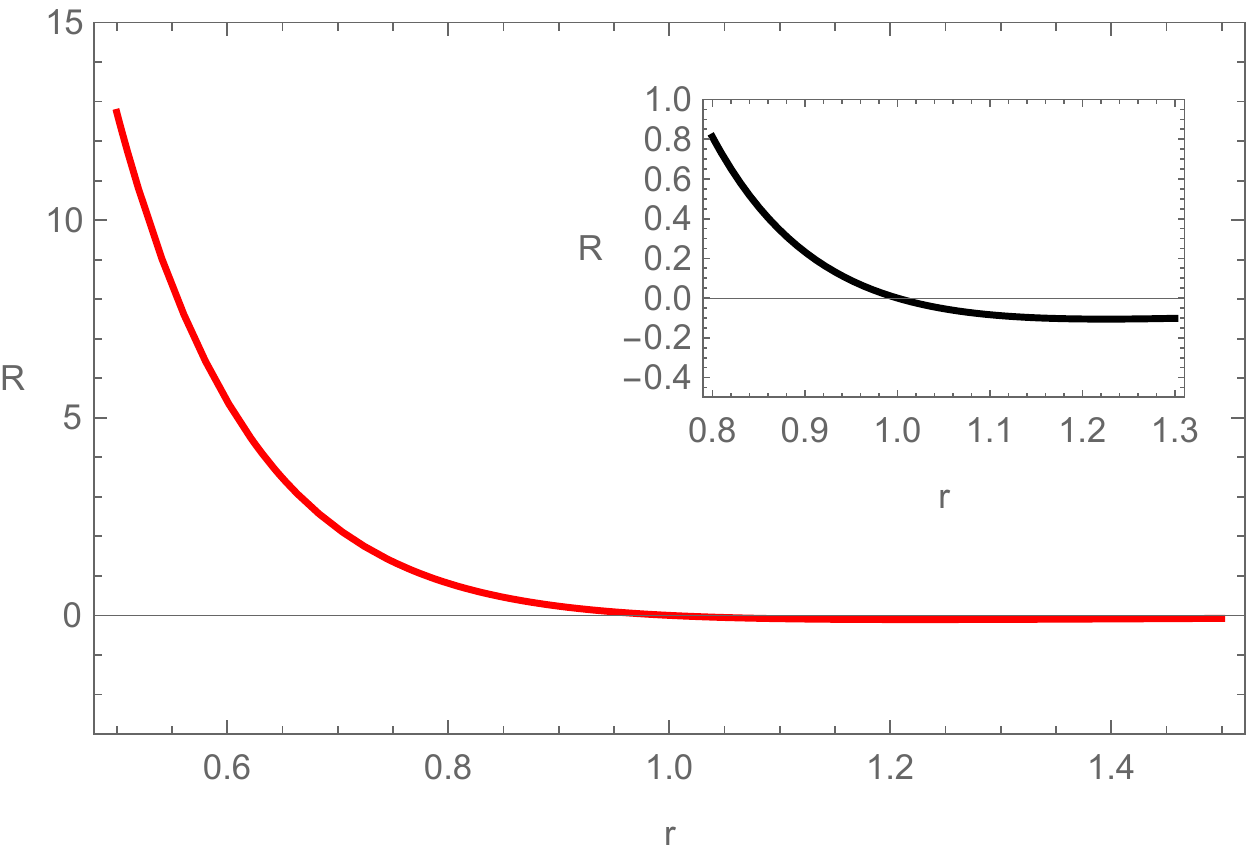}\hfill
\includegraphics[width=1\columnwidth,clip]{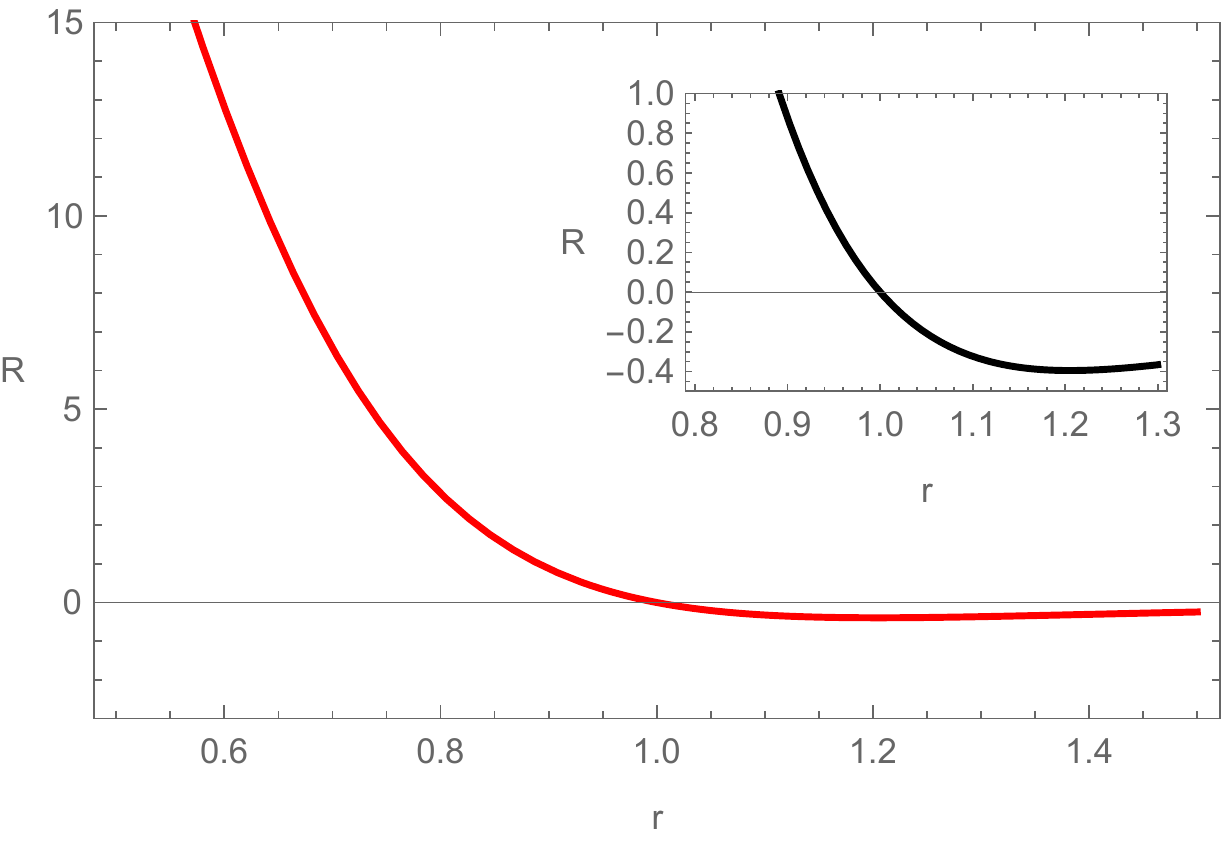}\hfill
\caption{Plots of the Ricci scalars  for Bardeen and Hayward spacetimes, left and right figures, respectively. We assume  unitary mass and $q=a=0.5$. The signs of $q$ and $a$ do not influence the curve shapes. In the Hayward case, the functional behavior at large radii is steeper than the Bardeen case, indicating stronger repulsion strength. At right, in the subfigures on top, we focus on the change of sign that approximately, with the aforementioned choice, occurs at the same point.}
\label{all_R_BH}
\end{figure*}

\begin{figure*}
 \centering
\includegraphics[width=1\columnwidth,clip]{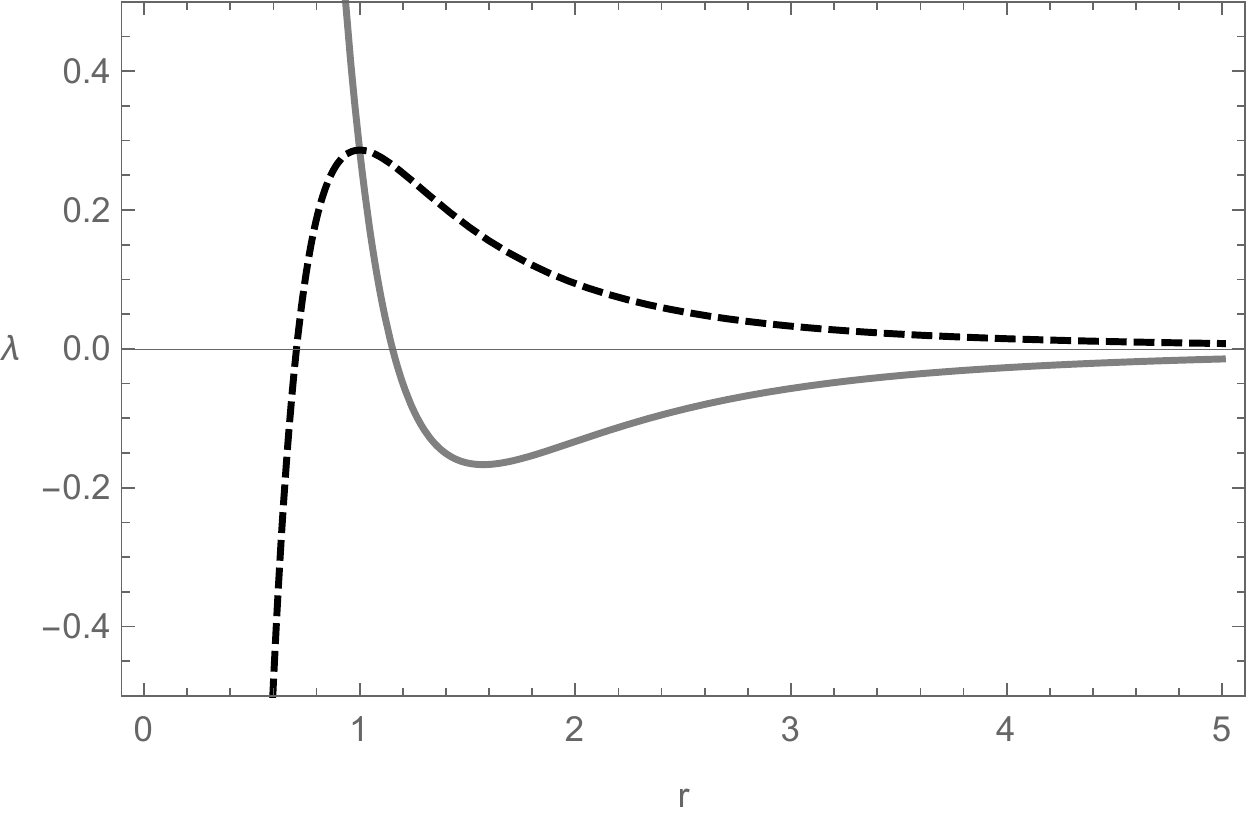}\hfill
\includegraphics[width=1\columnwidth,clip]{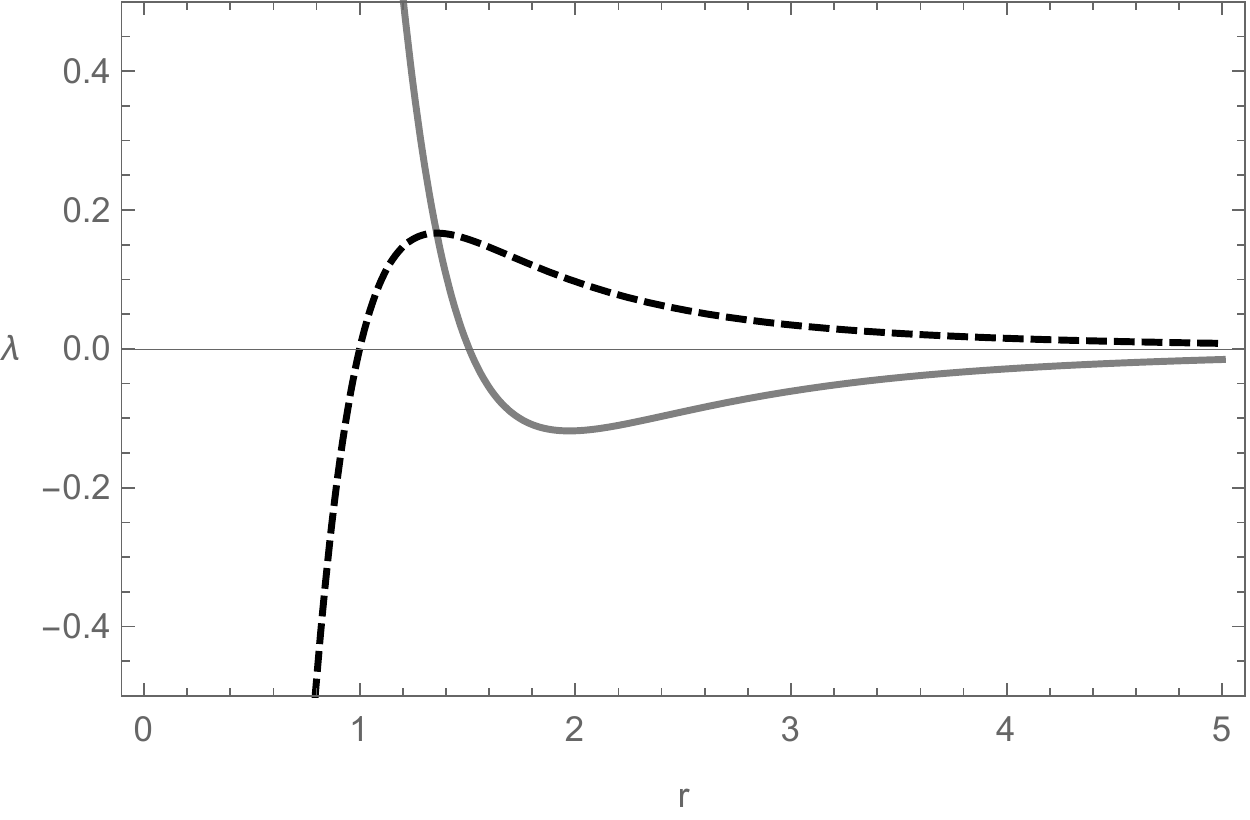}\hfill
\caption{Plots of the eigenvalues, $\lambda_1$ (gray, thick) and $\lambda_2$ (black, dashed) for Bardeen and Hayward spacetimes, left and right figures, respectively. We assume $M=1$ and $q=a=0.5$. The signs of $q$ and $a$ do not influence the curve shapes. }
\label{all_lambda_BH}
\end{figure*}

\subsection{Repulsive gravity in Hayward spacetime}

Almost forty years after the discovery of the Bardeen spacetime, Hayward introduced a new regular BH solution. The corresponding metric shares some similarities with the Bardeen metric, exhibiting, however, a flat center, where a specific matter energy-momentum tensor approaches a de Sitter contribution at the core, while vanishing for $r \to \infty$ \cite{2006PhRvL..96c1103H}. The lapse function for the Hayward regular BH is
\begin{equation}
f (r) =  1 - \frac{2 M r^2}{r^3 + 2 a^2}\,,
\end{equation}
where $M$ is the mass of the BH and $a$ is a constant, related to the de Sitter phase.

The curvature eigenvalues are
\begin{subequations}
\begin{align}
\lambda_1^{(H)} & = -\frac{2M(r^6 -14a^2r^3 + 4a^4)}{(r^3 + 2a^2)^3}\,, \\
\lambda_2^{(H)} & =  \lambda_3^{(H)} = - \lambda_5^{(H)} = - \lambda_6^{(H)} = \frac{M(r^3 - 4a^2)}{(r^3 + 2a^2)^2}\,, \\
\lambda_4^{(H)} & = \frac{2M}{r^3 + 2a^2}\,,
\end{align}
\end{subequations}
and the Ricci scalar reads
\begin{equation}
R^{(H)}=-24Ma^2\frac{(r^3-4 a^2)}{(2 a^2 + r^3)^3}\,.
\end{equation}

When approaching the source from infinity, the first extremum is located in $\lambda_1^{(H)}$ at $r^{(H)}_{rep}=3.13 a^{(2/3)}$, which determines the place of the repulsion onset. Furthermore,  $\lambda_1^{(H)}=0$ at $r=r^{(H)}_{dom}=2.39 a^{(2/3)}$, which indicates the region where repulsion dominates.  The remaining eigenvalues and the Ricci scalar do not provide additional information about repulsion in this spacetime.

The corresponding eigenvalues change signs at $r=(7 a^2 - 3 \sqrt{5} a^2)^\frac{1}{3}$, for $\lambda_1^{(H)}$, providing a
stationary point at $r_{rep}^\star=0$ and $r_{rep}^\star=(7a^2)^{\frac{1}{3}}$ while $r=(2a)^\frac{2}{3}$ and $r_{rep}^\star=0$ for $\lambda_{2;3;5;6}$. The point $r=0$ therefore represents both the regular shape, say the \emph{smoothness} of our solution, and the onset of acceleration.

To illustrate our approach, in Figs. \ref{all_R_BH} and \ref{all_lambda_BH}, we display the behaviors of the eigenvalues and Ricci scalar for given $M$, $q$ and $a$ values for both Bardeen and Hayward spacetimes\footnote{In the original formulation of the Hayward spacetime, $a^2\equiv M\Lambda^{-2}$, where $\Lambda$ is a de Sitter phase. For this reason, we  conclude that a de Sitter phase is on the core of the solution, \textit{i.e.}, for small $r$ it reduces to pure de Sitter. The corresponding plots are framed using unitary masses and arbitrary $\Lambda$ values, as emphasized in the corresponding captions. }.

\subsection{Repulsive gravity in Dymnikova spacetime}

In addition to the solutions mentioned earlier, the Dymnikova metric has been formulated based on the concept that the gravitational field of a BH can be described by a non-symmetric metric. This metric takes into account the existence of a non-zero energy-momentum tensor in the vicinity of the regular BH \cite{2004CQGra..21.4417D}. This approach enables a more precise depiction of the gravitational field generated by a BH and its impact on the surrounding spacetime, highlighting the possibility of incorporating non-linear electrodynamics effects into Einstein's equations.

The Dymnikova lapse function is given by
\begin{equation}
    f(r) = 1-\frac{2 M}{r} \frac{2}{\pi} \left[\arctan \left(\frac{r}{r_0}\right)
-\frac{rr_0}{r^2+r_0^2}\right]\,,
\end{equation}
where $r_0$ is the length scale, $r_0=\pi q^2/(8 M)$, $M$ is the total mass and $q$ is the charge that we will treat in analogy to the Bardeen solution, for the sake of simplicity.

The corresponding curvature eigenvalues are
\begin{subequations}
\begin{align}
\lambda_1^{(D)} & =  -\frac{4M}{\pi r^3 \arctan\left(  \frac{r^3}{ r_0( r^2 + r_0^2)}      \right) } , \\
\lambda_2^{(D)} & =  \lambda_3^{(D)} = - \lambda_5^{(D)} = - \lambda_6^{(D)} =
                      \frac{2M}{\pi r^3 \arctan\left(  \frac{r^3}{ r_0( r^2 + r_0^2)}      \right) }
 \ , \\
\lambda_4^{(D)} & =   \frac{4M}{\pi r^3 \arctan\left(  \frac{r^3}{ r_0( r^2 + r_0^2)}      \right) }  \ .
\end{align}
\end{subequations}

Moreover, we have for the Ricci scalar

\begin{equation}
R^{(D)}=\frac{2M}{\pi}\frac{r_0 r (r^4 + 4 r^2 r_0^2 - r_0^4) + (r^2 + r_0^2)^3 \arctan\left(\frac{r}{r_0}\right)}
{ r^2 (r^2 + r_0^2)^3}\,.
\end{equation}

From the above results, there is no evidence of sign change for the Dymnikova eigenvalues for any set of free constants $r_0$ and $M$. In addition, although the Ricci scalar is smooth at $r=0$, it does not change sign, indicating concordance between the approach that makes use of eigenvalues with that using first order curvature invariant.

\begin{figure*}
 \centering
\includegraphics[width=1\columnwidth,clip]{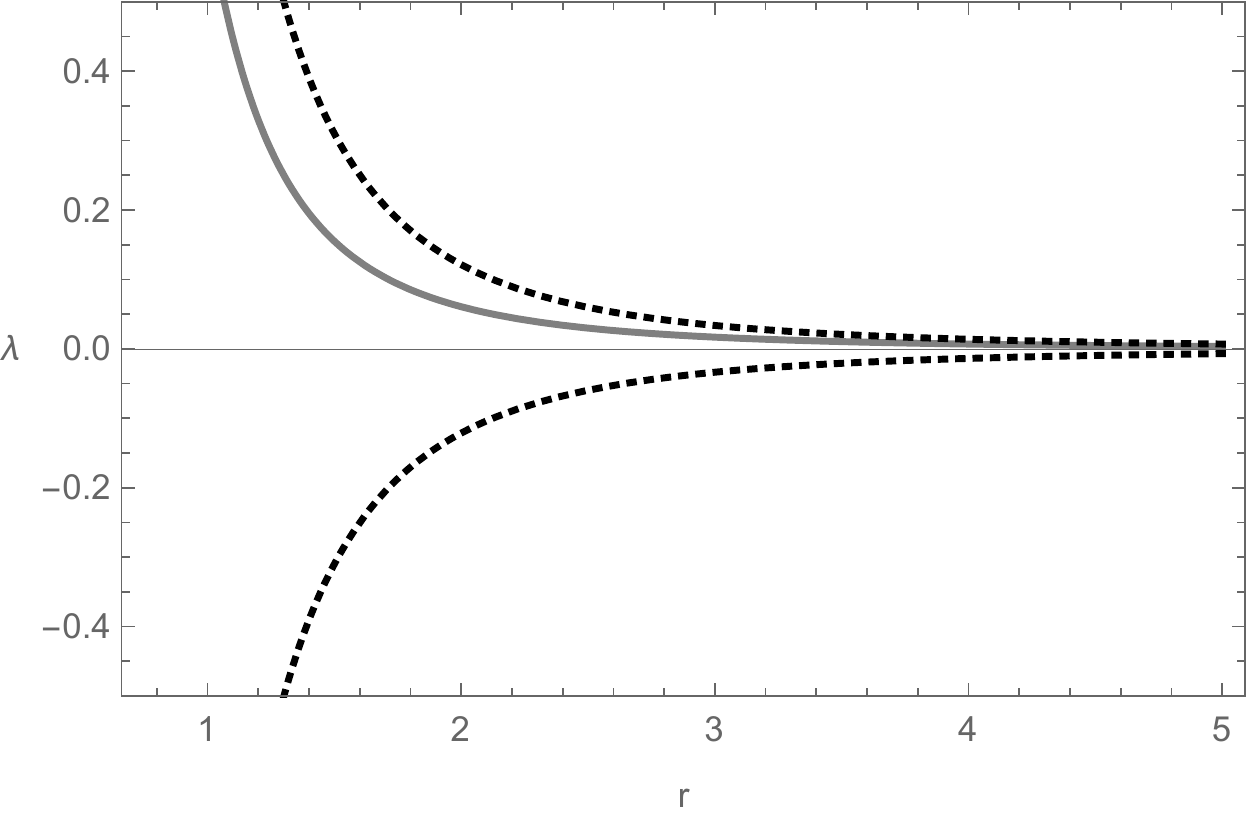}\hfill
\includegraphics[width=1\columnwidth,clip]{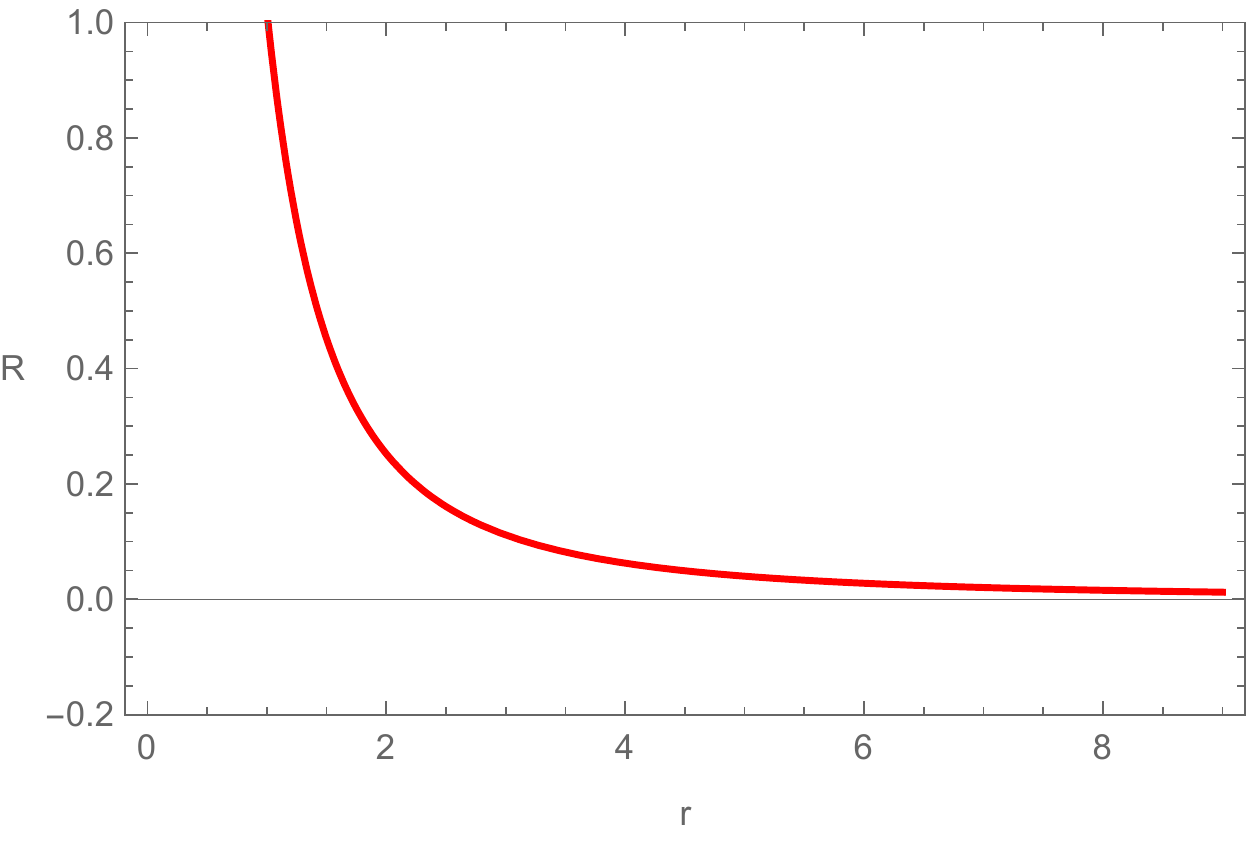}\hfill
\caption{Plots of the eigenvalues, $\lambda_1$ (gray, thick) and $\lambda_2$ (black, dashed) and of the Ricci scalar (right figure) for the Dymnikova  spacetime. Conventional unitary $r_0$ has been used. No regions of repulsion have been found.}
\label{all_lambda}
\end{figure*}

\section{Theoretical discussion}\label{sezione4}

The regions of repulsive gravity within the aforementioned solutions can provide insights into the "completeness" of regular BHs. Indeed, it is worth noting that conventional BHs, \textit{i.e.}, those exhibiting singularities, inherently possess repulsive regions that arise from the interplay between the mass generating the BH and the free parameters of the specific solution \cite{Luongo:2014qoa}.

In our analysis of regular BHs, we notice that the signs of $q$ and $a$ do not influence $R$ to change its sign as well as the sign of $r_0$. The effects become evident changing $\lambda_1$ into $\lambda_4$, modifying the sign of $\lambda_2$, only for $q$ and $a$. In fact, only in Bardeen and Hayward spacetimes there are regions of repulsion. In Dymnikova spacetime, there is no repulsive gravity, as confirmed by the study of the curvature eigenvalues and the Ricci scalar.

\begin{figure*}
 \centering
\includegraphics[width=1\columnwidth,clip]{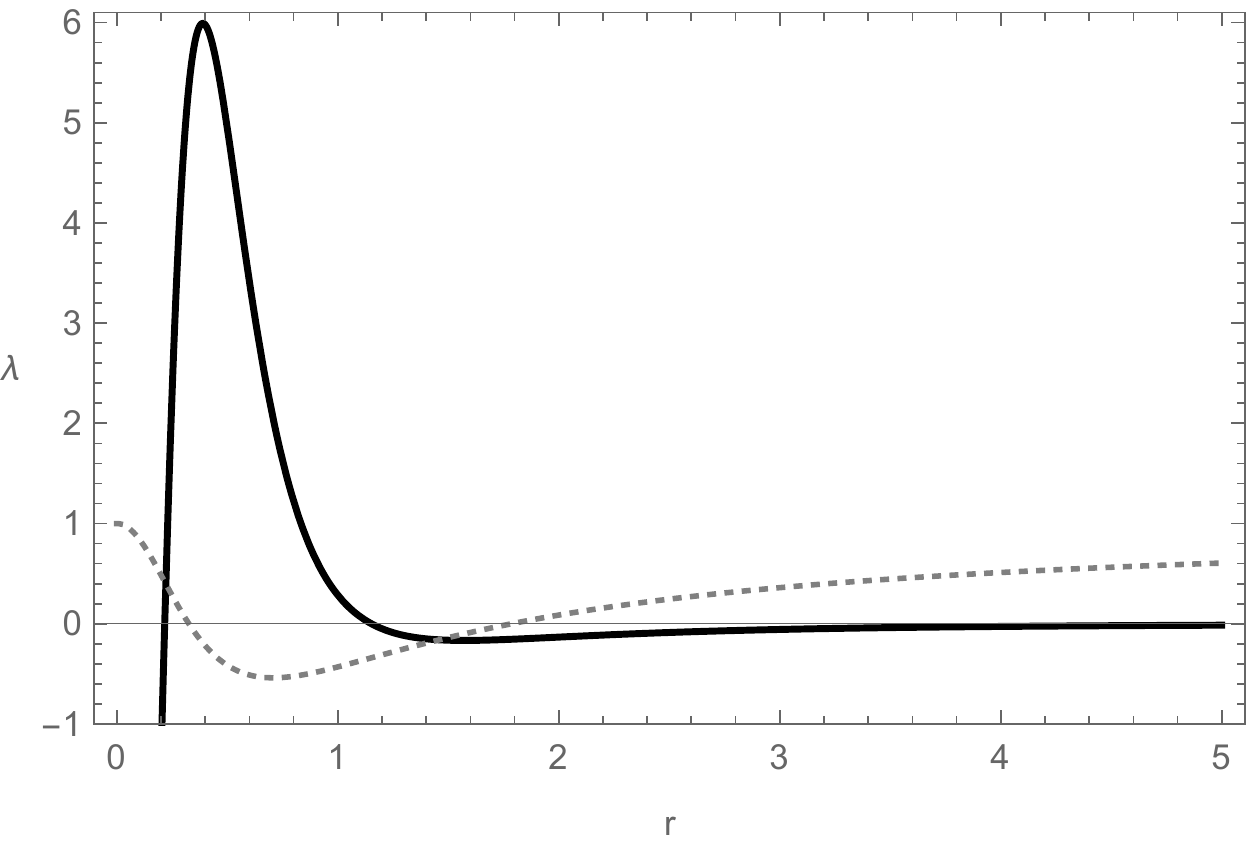}
\includegraphics[width=1\columnwidth,clip]{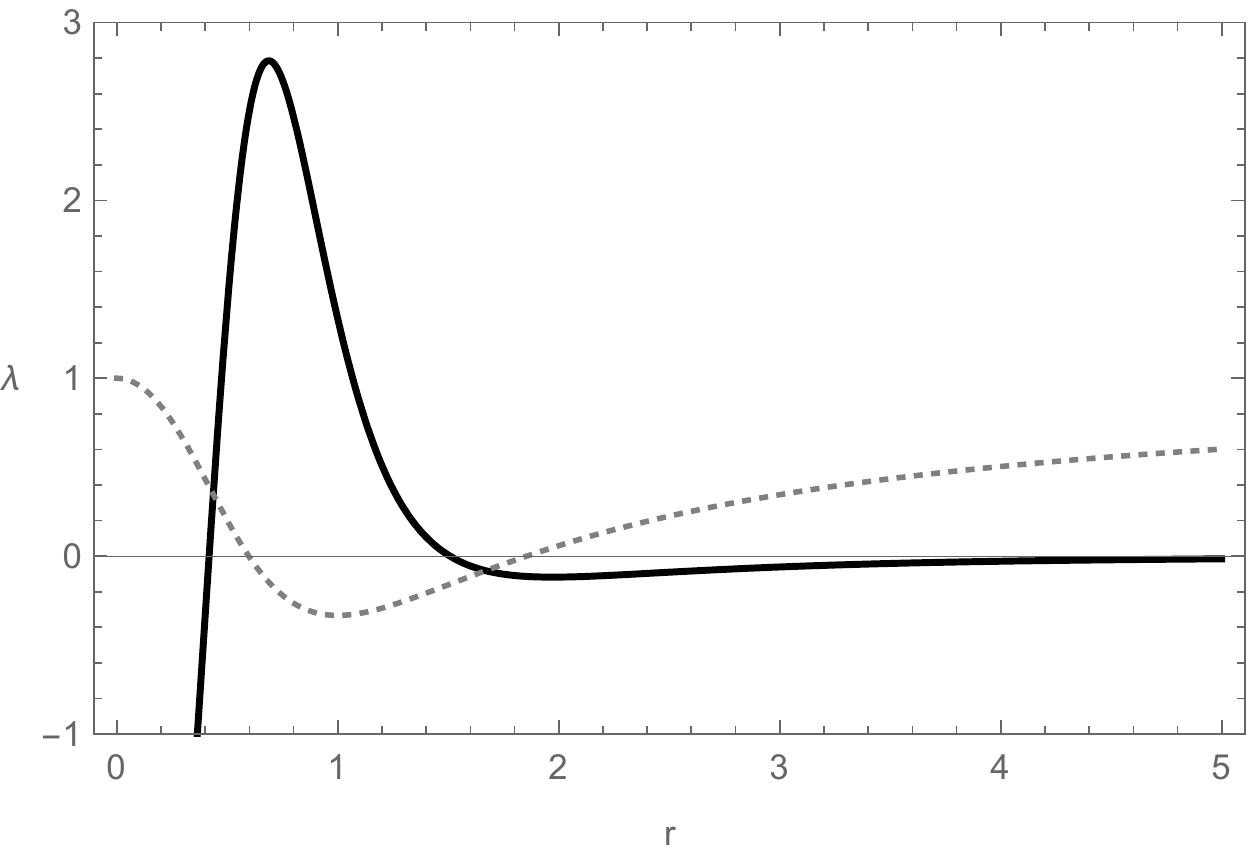}
\caption{Plots of the eigenvalues $\lambda_1^{(B)}$ and $\lambda_1^{(H)}$ (solid curves, called generically $\lambda$ on the $y$-axis) and the lapse functions (dotted curves) of the Bardeen  (left figure) and Hayward (right figure) metrics, respectively. Here, we choose the free parameters as $M=1$ and $q=a=1/2$. The zeros of the eigenvalue determine the points where repulsive gravity becomes dominant. The zeros of the lapse function correspond to the horizons.}
\label{fighor}
\end{figure*}

Moreover, as in the case of singular BHs \cite{Luongo:2014qoa},  repulsive gravity in regular BHs is located in regions close to the horizons. As an illustrative example, we portray this result in the case of the Bardeen and Hayward metrics in Fig. \ref{fighor}, where we plot the eigenvalues $\lambda_1^{(B)}$ and $\lambda_1^{(H)}$ (solid curves, with $y$-axis called generically $\lambda$), which, as shown above, determine the regions of repulsion, and the lapse functions (dotted curves), which determine the locations of the event horizons. For instance, in the Bardeen case we see that the onset of repulsion occurs at $r_{rep}=1.57$, the dominance radius is located at $r_{dom}=1.15$, whereas the exterior horizon is located at $r_{hor}=1.78$. Consequently, the region where repulsive effects become evident is located inside the exterior horizon. This result is important for the description of the gravitational field of compact objects by means of regular BH solutions. Indeed, in realistic situations, the horizon is located close to the center and far below the surface of compact objects, implying that repulsion effects are difficult, if not impossible, to be observed outside the surface of compact objects.

The case of the Dymnikova regular solution is different, since as stated above, there are no regions of repulsive gravity. On the other hand, it is important to stress that the  Dymnikova metric does not pathologically pass experimental bounds, e.g. see Ref. \cite{Boshkayev:2023rhr}. So, one can wonder whether exhibiting repulsive gravity could represent a \emph{necessary condition} to guarantee that a given regular spacetime, or more broadly any possible spacetime, can be used to describe the gravitational field of compact objects.

There remains the case of $M\rightarrow-M$ for all the solutions\footnote{Consider the Schwarzschild solution and suppose  that repulsive gravity exists. Then, without analyzing eigenvalues, the only way to obtain repulsive gravity, in view of the fact that the Ricci scalar vanishes, is to invert the sign of the mass. Negative masses imply the existence of \emph{exotic matter}, whose existence is theoretically plausible from wormhole physics, see e.g. \cite{Visser:1995cc,Morris:1988cz,Capozziello:2020zbx}, and cosmological dynamics, see e.g. \cite{Luongo:2012dv,Luongo:2014nld,Aviles:2016wel,Dunsby:2016lkw}.}. Even though this could be, in principle, possible for the Bardeen and Hayward cases, the effect of modifying the mass sign implies only to invert the intervals found. On the other hand, in the Dymnikova spacetime, this change does not modify the physics of the corresponding spacetime.

These differences of our findings, or, naively, the strange behavior of regular BHs, can be compared with geometric issues related to the Dymnikova metric \cite{Zhou:2022yio} and with extended versions of Hayward regular BHs, where higher orders invariants appear singular \cite{Giacchini:2021pmr}, suggesting the Bardeen BH be solely safe from singularities at higher orders. In addition, the Hayward BH would be filtered out by the path integral approach to quantum gravity, while the Bardeen BH would survive. Last but not least, both Bardeen and Hayward spacetimes show islands that degenerate with the standard Schwarzschild solution \cite{Luongo:2023jyz}, appearing similar in the corresponding thermodynamics \cite{Myung:2007av}.

Consequently, we \emph{cannot conclude} that the effects of regularity at all curvature levels are related to repulsion effects, but rather that, at least at  the first order, this may happen, as above stated. On the contrary, the Dymnikova spacetime still persists with issues as a consequence of the above outcomes.

\subsection{Comparing with singular solutions}

In order to compare our regular BHs with standard solutions,  we here work out the regions of repulsive gravity in
the  Reissner-Nordstr\"{o}m solution with a de Sitter phase, whose lapse function can be written under the form
\begin{equation}\label{generalissima}
    f(r)\equiv 1-\frac{2M}{r}+\frac{q^2}{r^2}-\Lambda r^2\,.
\end{equation}
We conventionally chose Eq. \eqref{generalissima} since the solutions for Reissner-Nordstr\"{o}m and Schwarzschild-de Sitter emerge as one takes $\Lambda\rightarrow0$ and $q\rightarrow0$, respectively\footnote{We purposely choose the same topological charge and vacuum energy contribution prompted in the Bardeen and Hayward spacetimes, respectively.}.

Using the metric function \eqref{generalissima}, we obtain the following eigenvalues
\begin{align}\label{altri}
&\lambda_1^\theta=-\frac{1}{r^3}\Big[2M-\frac{3Q^2}{r}\Big]-\Lambda\,,\nonumber\\
&\lambda_2^\theta=-\lambda_1\,,\nonumber\\
&\lambda_3^\theta=\frac{1}{r^3}\Big[M-\frac{Q^2}{r}\Big]-\Lambda\,,\\
&\lambda_4^\theta=-\lambda_5=\lambda_6=-\lambda_3\,,\nonumber
\end{align}
where $\theta\equiv \left\{SdS; RN\right\}$, reducing to the Reissner-Nordstr\"{o}m, $\lambda^{RN}$, and Schwarzschild-de Sitter, $\lambda^{SdS}$, cases as $\lambda(\Lambda=0)\equiv \lambda^{RN}$ and $\lambda(q=0)\equiv \lambda^{SdS}$,  respectively. Using the results presented in Eqs. \eqref{altri}, we compare $\lambda^{RN}$ and $\lambda^{SdS}$ with the eigenvalues of the regular BHs analyzed above.

At a first analysis, the most crucial property of regular metrics is that the onset of repulsive gravity \emph{does not depend on linear combinations between matter and charge or vacuum energy}. In other words, the value of mass does not influence  the behavior of the regular eigenvalues, contrary to what happens in singular BH solutions.

Although it may appear that the effect of the singularity would be to  add a new extra parameter into the computation of repulsive regions, we prefer to rephrase this result as
showing that the eigenvalues of regular solutions do not depend on the mass of the object.

Specifically, we calculate the masses of the BHs that replicate the results obtained from regular BHs. This is achieved by determining the zeros of each eigenvalue and equating them to the zeros of the eigenvalues of the Reissner-Nordstr\"{o}m and Schwarzschild-de Sitter solutions. Consequently, we establish that there always exist specific masses that mimic the behaviors of regular solutions. The numerical findings are summarized in Table \ref{Table}. Based on these results, we put forward the conjecture that there is an inherent incompleteness in regular BHs, or alternatively, that singular BHs are more general than their regular counterparts. As a consequence, the effects observed in regular solutions are somehow replicated in singular BHs. In other words, the effects of repulsive gravity are analogous once the value of masses are imposed as in Tab. \ref{Table}. Last but not least, from Tab. \ref{Table} we see that the presence of opposite radii implies that the sign of $\Lambda$ and $q$ might be opposite (discord) and not precisely defined.

\section{Final outlooks and perspectives}\label{sezione5}

In this work, we investigated regions of repulsive gravity in three regular BH solutions,  namely, the Bardeen, Hayward, and Dymnikova spacetimes. To do so, we focused on first order geometric invariants, involving the Ricci scalar and the eigenvalues of the Riemann curvature tensor. The first choice is physically motivated as one expects the Ricci scalar to be sensible to the change $M\rightarrow-M$, implying regions of repulsion. The second
method is motivated by the eigenvalue property of vanishing asymptotically (in asymptotically flat spacetimes) and to increase as the source of gravity is approached, providing a radius at which the passage from attractive to repulsive gravity might occur.

We found the corresponding locations of the onset and the regions of repulsion, showing that the Bardeen and Hayward solutions contain regions of repulsive gravity, whereas the Dymnikova spacetime is free of repulsion. Remarkably, we also  showed that repulsive gravity is found without linearly combining the mass term with topological charge and vacuum energy, for the Bardeen and Hayward spacetimes. As the mass term is supposed to generate the corresponding solution, the fact that repulsive regions do not depend on linear combinations with it appears nonphysical, leading to an intrinsic pathology of those models, that we conjectured as \emph{incompleteness} of the employed regular solutions. To this end, by using the first order invariants, we thus compared our findings with the Reissner-Nordstr\"om and Schwarzschild-de Sitter spacetimes. We found values of masses that permit to emulate the same repulsive effect of Bardeen and Hayward metrics, showing that, at the level of repulsive effects, those regular solutions are somehow less general than singular BH solutions. We compared our findings with previous results in the literature, confirming theoretical inconsistencies that have been raised against the use of regular metrics.

Even though it is evident that those solutions may exhibit strange behaviors at the level of repulsive regions, extensions of such spacetimes can be investigated, for future developments, requiring the presence of the mass combinations in the expressions of the repulsive radii. Hence, under this requirement, one can search for new regular solutions that satisfy this condition. Finally, it would be interesting to investigate other classes of regular solutions, including additional gravitational effects such as quadrupole, rotation, etc., to check the goodness of such objects by looking at the predicted sign of gravity.

\section*{Acknowledgements}
OL acknowledges Roberto Giamb\`o for fruitful discussions on the subject of this paper. The work of OL is was partially financed by the Ministry of Education and Science of the Republic of Kazakhstan, Grant: IRN AP19680128. The work of HQ was partially supported  by UNAM-DGAPA-PAPIIT, Grant No. 114520, and CONACYT-Mexico, Grant No. A1-S-31269.

\clearpage
\appendix

\section*{Bivector representation of the curvature tensor}

In the bivector representation, the Riemann curvature tensor can be expressed as
the 6$\times$6 matrix given in Eq. \eqref{eq: CurvatureTensor}, where the explicit form of the matrices is

\begin{equation}
{\bf L} =\left(
\begin{array}{ccc}
{\bf R}_{14}  & 	{\bf R}_{15}  & {\bf R}_{16} \\
{\bf R}_{15} - \kappa T_{03} &  {\bf R}_{25} &  {\bf R}_{26}  \\
{\bf R}_{16} + \kappa T_{02}  &   \quad {\bf R}_{26}  - \kappa  T_{01} & \quad - {\bf R}_{14} -	{\bf R}_{25}   \\
\end{array}
\right),
\end{equation}

\noindent whereas ${\bf M}_1$ and  ${\bf M}_2$ are $3\times 3$ symmetric matrices of the form

\begin{equation}
{\bf M}_1=\left(
\begin{array}{ccc}
{\bf R}_{11} &  \quad	{\bf R}_{12} & {\bf R}_{13} \\
{\bf R}_{12} & \quad {\bf R}_{22} &  {\bf R}_{23} \\
{\bf R}_{13} & \quad   {\bf R}_{23} &  \quad - {\bf R}_{11}   -	{\bf R}_{22}  {+} \kappa \left(\frac{T}{2} +T_{00}\right)  \\
\end{array}
\right),
\end{equation}

\noindent and

\begin{widetext}
\begin{equation}
{\bf M}_2 =
{   \left( \\
	\begin{array}{ccc}
	-{\bf R}_{11} + \kappa \left(\frac{T}{2} +T_{00}-T_{11} \right)   &  {-} { \bf R}_{12} - \kappa T_{12}   & - {\bf R}_{13} - \kappa T_{13} \\
	{-} { \bf R}_{12} - \kappa T_{12}   &   -{\bf R}_{22} + \kappa \left(\frac{T}{2} +T_{00}-T_{22} \right)   &  - {\bf R}_{23} - \kappa T_{23} \\
	- {\bf R}_{13} - \kappa T_{13}    &    - {\bf R}_{23} - \kappa T_{23}   &   {\bf R}_{11} +	{\bf R}_{22}  {-} \kappa T_{33}   \\
	\end{array}
	\right)   }\,,
\end{equation}
\end{widetext}
where $\kappa\equiv 8\pi$.
Notice that the traces of the matrices  satisfy the properties $
{\rm Tr}({\bf M_1}) = \kappa \left(\frac{T}{2} +T_{00}\right)$, ${\rm Tr}({\bf M_2}) =  \kappa T_{00}$ and, finally, the trace of the Riemann curvature matrix  becomes
${\rm Tr}({\bf R_{AB}}) = \kappa \left(\frac{T}{2} +2 T_{00}\right)$, where  $T=\eta^{ab}T_{ab}$.

\end{document}